\input{aipcheck}

\documentclass[
    ,numbers            
  ]
  {aipproc}

\layoutstyle{6x9}

\renewcommand\XFMtitleblock{%
 \XFMtitle
 \let\XFMoldpar\par
 \def\par{\XFMoldpar\def\par{\space
     for the VERITAS Collaboration\footnote{\protect\url{http://veritas.sao.arizona.edu/}}\XFMoldpar}}%
  \XFMauthors
  \let\par\XFMoldpar
  \XFMaddresses
  \XFMabstract
  \vspace{5pt}%
  \XFMkeywords
  \XFMclassification
}

\begin{document}

\title{Observation of binary systems at very-high energies with VERITAS}

\classification{95.85.Pw;98.38.Fs;97.80.Jp}
\keywords      {gamma-rays, binary, LS I +61 303, HESS J0632+057, VERITAS}

\author{G.Maier}{
  address={DESY, Platanenallee 6, 15738 Zeuthen
}}

\begin{abstract}
Non-thermal variable emission from radio to very-high energy gamma rays (VHE; $>$100 GeV) are the prime characteristics of gamma-ray binaries. The underlying physical processes leading to the observed VHE emission are not well understood, as even the most basic features of these systems are under dispute (microquasar model vs shocked pulsar wind model). VHE binaries can be difficult to observe, some have orbital periods of the order of years (e.g. HESS J0632+057 or PSR B1259-69) or show irregular emission patterns as observed in LS I +61 303.

We present here new VERITAS observations of the binary systems  LS I +61 303 and HESS J0632+057 carried out with higher sensitivity and more dense temporal coverage than previous observations. The gamma-ray results and their astrophysical implications are discussed in the context of contemporaneous observations with Swift XRT and Fermi LAT at X-ray and gamma-ray energies. \end{abstract}

\maketitle


\section{Introduction}

Gamma-ray binaries constitute a small class of Galactic objects with currently less than ten members detected.
Their common feature is variable non-thermal emission across all wavelengths, in most cases modulated 
by the orbital motion as both the massive star and the compact object orbit the center-of-mass.
Why only a handful of the several hundreds of binaries in the Milky Way shine in gamma rays is unclear, as are the underlying particle acceleration and gamma-ray emission mechanisms.
The modeling of these system is difficult due to many unknowns (e.g. geometry of the orbit, nature of the compact object) and many time-dependent variables (e.g. wind-jet interaction, stellar disks, temporary accretion disk, jet interaction with the circumstellar environment).
Among the possible gamma-ray production mechanisms are the interaction of relativistic particles from a pulsar wind with  UV photons of the massive star or with its dense equatorial wind. Alternatively, the acceleration of particles in a jet powered by accretion of mass onto the black hole or neutron star from the companion star \cite{Mirabel-2012} can produce a high-energy particle population.

New gamma-ray observations with greater sensitivity and more dense temporal coverage over many orbital periods with simultaneous coverage over several wavebands  can offer new insights.
Observations of gamma-ray binaries are challenging: 
the emission can be highly variable (sometimes seemingly unpredictable as in LS I +61 303 \cite{Acciari-2011}),
the orbital period can be close to the length of typical observing periods (e.g. LS I +61 303 with an orbital period of 26.5 days, very close to the cycle of the moon), or very long (e.g. 3.4 years in PSR 1259-63).
We present in the following new observation of the binaries LS I +61 303 and HESS J0632+057 at very-high energies ($>100$ GeV) with VERITAS, a system of four ground-based imaging atmospheric Cherenkov telescopes located in southern Arizona
(see \cite{Galante-2012} for details).

\section{LS I +61 303}

\begin{figure}
  \includegraphics[width=0.77\linewidth]{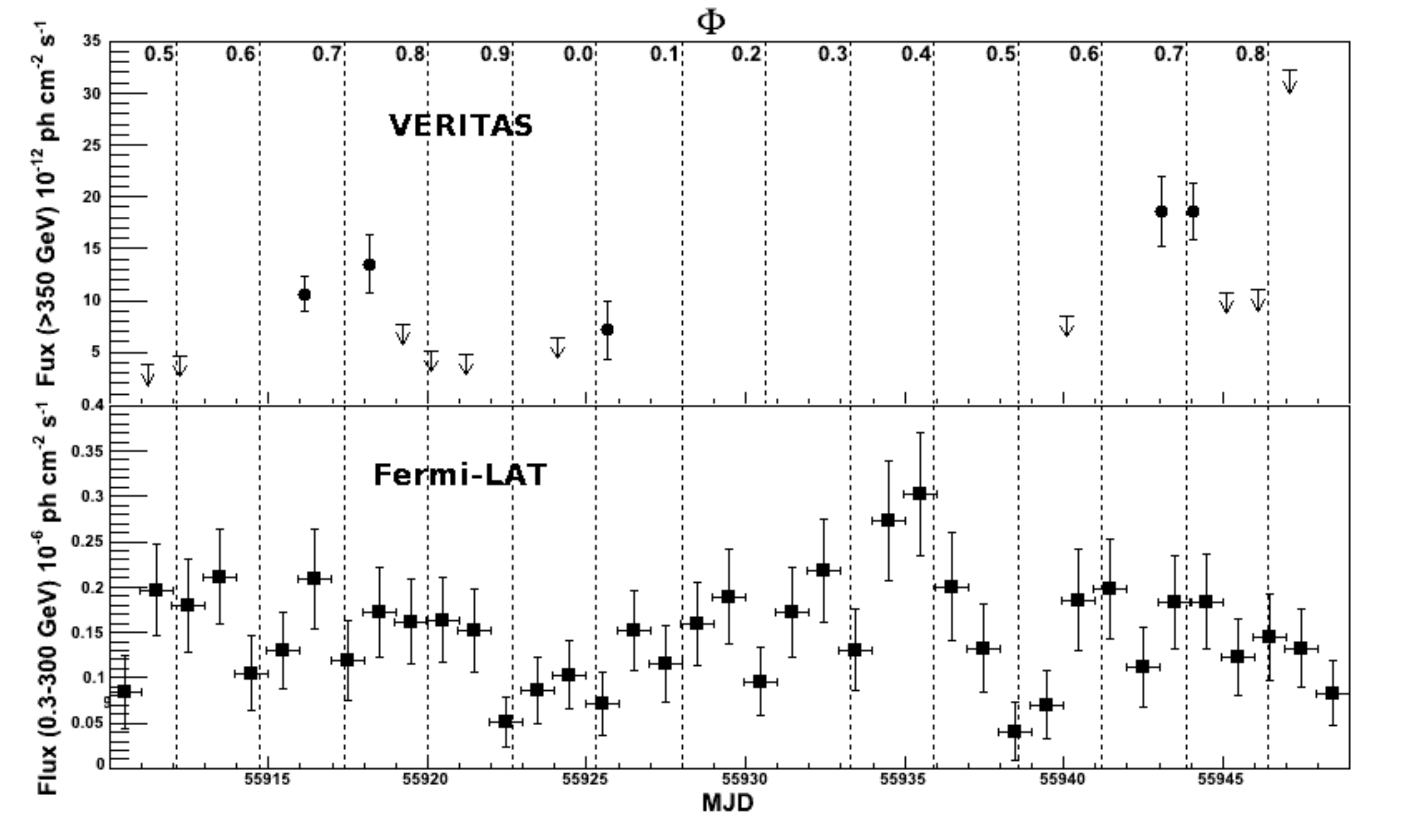}
  \caption{High-energy and very-high energy observations from Dec 2011 to Jan 2012 of LS I +61 303 with Fermi LAT (bottom) and VERITAS (top). 
  Upper limits on the  flux (95\% probability) are shown for VERITAS data points with significances less than $3\sigma$.
  Orbital phases are indicated on the top of the figure using an  orbital period of $P = 26.4960 \pm 0.0028$ days and zero orbital phase of $T_0 =\mathrm{ HJD}  \ 2, 443, 366.775$ \cite{Gregory-2002}. }
\end{figure}

The gamma-ray binary LS I +61 303 consist of a Be star and a compact object of unknown nature.
The underlying acceleration and gamma-ray emission scenario is actively discussed in literature, with several inconclusive observations.
As an example, the radio observations with VLBA in 2006 were first interpreted as proof for the pulsar wind nature of LS I + 61 303 \cite{Dhawan-2006}, while a recent reanalysis of the data revealed a double sided jet structure \cite{Zimmermann-2012}. 
LS I +61 303 has been detected in VHE by MAGIC and VERITAS mostly at phases around 0.6-0.7 (close to apastron) \cite{Albert-2006, Acciari-2008}, but also occasionally close to periastron \cite{Acciari-2011}. 
The binary has not been detected close to apastron despite deep observations by VERITAS in 2008 and 2009.

Fig 1 shows the results from observations of LS I +61 303 around apastron for two orbits  in Dec 2011 to Jan 2012  with VERITAS and Fermi LAT.
The object is clearly detected in 25 hours of high elevation observations with 12 standard deviations above background. 
The measurements of VERITAS reveal time variability on timescales of the order of 1 day 
(integral flux $>350$ GeV for MJD 55944: $(1.9\pm 0.3)\cdot 10^{-11}$ cm$^{-2}$s$^{-1}$, detection significance $9\sigma$ in 100 min;  
upper limit on integral flux $>350$ GeV for MJD 55945: $< 1.07\cdot 10^{-11}$ cm$^{-2}$s$^{-1}$ (95\% probability), detection significance $2.5\sigma$ in 90 min).  
The detection of  fast variability at VHE certainly adds another level of complexity to the interpretation of the non-detection of the system in 2008 and 2009.
It is unclear at this point if the orbit-to-orbit variability observed over the past few years is due to intrinsic multi-year modulation or if the observations
simply miss the possibly very short episodes of VHE emission.

\section{HESS J0632+057}

\begin{figure}
  \includegraphics[width=0.77\linewidth]{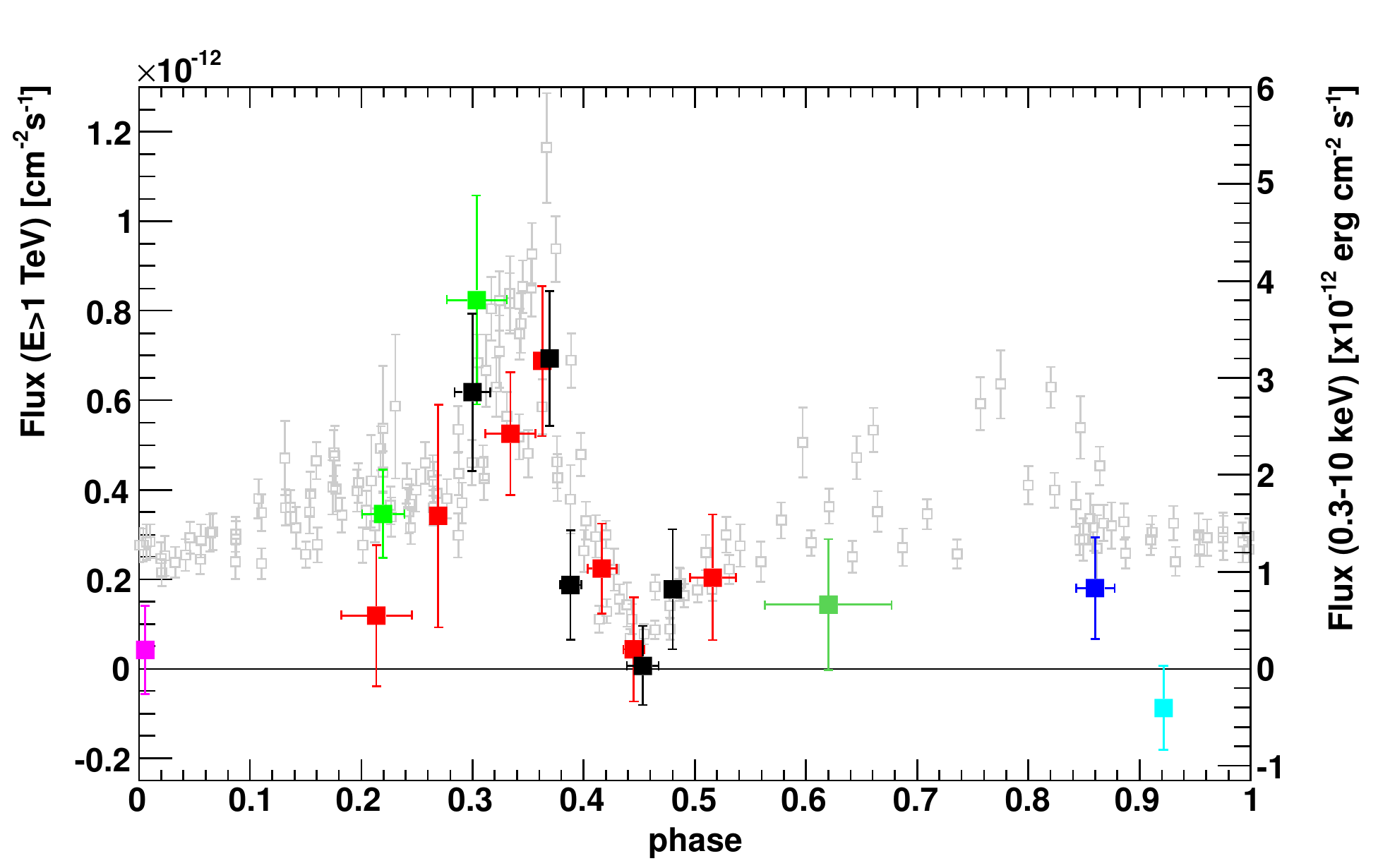}
  \caption{
  Integral gamma-ray fluxes above 1 TeV of HESS J0632+057 from VERITAS  (filled markers) and Swift XRT X-ray measurements (0.3-10 keV) folded with the orbital period of 315 days (see text). 
  The colors indicate the different observing periods of VERITAS: Nov 2011-Feb 2012 (black), Dec 2010-April 2011 (red), Feb 2010-March 2010 (light green), Oct 2010 (dark blue), Jan 2009 (pink), Dec 2008 (light blue) and Dec 2006-Jan 2007 (dark green).
  }
\end{figure}

Gamma-ray emission from HESS J0632+057 above 400 GeV was serendipitously discovered during observations of the Monoceros Loop Supernova Remnant in 2004/2005 \cite{Aharonian-2007}. 
Follow-up observations by VERITAS revealed a variable gamma-ray source \cite{Acciari-2009}, coincident with the hard X-ray source XMMU J063259.3+054801 \cite{Hinton-2009} and the Be star MWC 148.
Multi-year X-ray observations with Swift XRT revealed unambiguously the binary nature of HESS J0632+057  \cite{Falcone-2010, Bongiorno-2011}.
The orbital solution from optical spectroscopy of the optical counterpart MWC 148 of the gamma-ray source supports an eccentricity of $e=0.83\pm0.08$, with the maximum in the X-ray emission about 0.3 phases after periastron \cite{Casares-2012}. 

The analysis of the enlarged Swift XRT data set (from MJD 54857 to MJD 55972) using z-correlated discrete correlation functions results in an orbital period for HESS J0632+057 of $315^{+6}_{-4}$ days (see \cite{Bongiorno-2011} for details on the X-ray observations).
The  phase-folded X-ray light curve (Fig 2) shows several distinct features which are not unusual for binaries: a maximum between phases 0.3-0.4 followed by a dip at 0.45 (due to shadowing or suppression of the wind-wind interaction?) and a possible second maximum at phases 0.75.
The gamma-ray fluxes from 144 hours of VERITAS data from several orbital periods of HESS J0632+057 are shown in Fig 2. 
The data from 2012 are presented for the first time, the deep exposure of 34 h  resulted in a detection significance of $9.8\sigma$.
The gamma-ray light curve follows the X-ray light curve, with a repeated detection at phases 0.3-0.4.
The time-lag between X-ray and gamma-ray emission is consistent with zero ($-4^{+7}_{-14}$ days).
A detailed analyses of the long-term observations of HESS J0632+057 with VERITAS, HESS and Swift will be presented in an upcoming publication \cite{Acciari-2012, Bordas-2012}.

\section{Conclusions}

The deep observation over several years of the two gamma-ray binaries LS I +61 303 and HESS J0632+057 with VERITAS revealed very different emission patterns: while LS I +61 303 seems to become more puzzling with ongoing observations, shows HESS J0632+057 a  predictable light curve with maximum VHE emission simultaneous with the maximum in X-ray emission.
The enhanced sensitivity of VERITAS after its focal plane upgrade in 2012 will allow improved observations of these binaries in the upcoming years.


\vspace{0.3cm}

{\small
{\bf ACKNOWLEDGMENTS} This research is supported by grants from the US Department
of Energy, the US National Science Foundation, and the Smithsonian
Institution, by NSERC in Canada, by Science Foundation
Ireland, and by STFC in the UK.
We acknowledge the
excellent work of the technical support staff at the FLWO and
the collaborating institutions in the construction and operation
of the instrument.
GM acknowledges support through the Young Investigators Program of the Helmholtz Association.
This work made use of data supplied by the UK Swift Science Data Centre at the University of Leicester.
}

\end{document}